\documentclass{article}

\usepackage{PRIMEarxiv}

\usepackage[utf8]{inputenc} 
\usepackage[T1]{fontenc}    
\usepackage{hyperref}       
\usepackage{url}            
\usepackage{booktabs}       
\usepackage{amsfonts}       
\usepackage{nicefrac}       
\usepackage{microtype}      
\usepackage{lipsum}
\usepackage{fancyhdr}       
\usepackage{graphicx}       
\graphicspath{{media/}}     
\usepackage{amsmath}
\usepackage{makecell} 
\usepackage{threeparttable}
\usepackage{amssymb}
\usepackage{framed,multirow}
\usepackage{caption}
\usepackage{color}
\usepackage{rotating}
\title{A Multi-Scale Spatial Transformer U-Net for Simultaneously Automatic Reorientation and Segmentation of 3D Nuclear Cardiac Images
\thanks{\textit{Corresponding author}: 
Wentao Zhu \\
\qquad $^ \dagger$ Both authors contribute equally to this work} 
}
\author{
  Yangfan Ni $^{\dagger}$, Duo Zhang $^{\dagger}$, Gege Ma \\
  Research Center for Human-Machine Augmented Intelligence \\
  Zhejiang Lab \\
  Hangzhou\\
  \texttt{niyangfan@zhejianglab, zhang\_duo@foxmail.com} \\
  \texttt{gegema@zhejianglab.com} \\
  \And
  Lijun Lu \\
  School of Biomedical Engineering and Guangdong Provincial Key Laboratory of Medical Image Processing \\
  Southern Medical University \\
  Guangzhou\\
  \texttt{ljlubme@gmail.com} \\
  \And
  Zhongke Huang \\
  Department of Nuclear Medicine \\
  Sir Run Run Shaw Hospital \\
  Hangzhou\\
  \texttt{3200021@zju.edu.com} \\
  \And
  Wentao Zhu \\
  Research Center for Human-Machine Augmented Intelligence \\
  Zhejiang Lab \\
  Hangzhou\\
  \texttt{wentao.zhu@zhejianglab.com} \\
}

\begin{document}
\maketitle

\begin{abstract}
Accurate reorientation and segmentation of the left ventricular (LV) is essential for the quantitative analysis of myocardial perfusion imaging (MPI), in which one critical step is to reorient the reconstructed transaxial nuclear cardiac images into standard short-axis slices for subsequent image processing. Small-scale LV myocardium (LV-MY) region detection and the diverse cardiac structures of individual patients pose challenges to LV segmentation operation. To mitigate these issues, we propose an end-to-end model, named as multi-scale spatial transformer UNet (MS-ST-UNet), that involves the multi-scale spatial transformer network (MSSTN) and multi-scale UNet (MSUNet) modules to perform simultaneous reorientation and segmentation of LV region from nuclear cardiac images. The proposed method is trained and tested using two different nuclear cardiac image modalities: $^{13}$N-ammonia PET and $^{99m}$Tc-sestamibi SPECT. We use a multi-scale strategy to generate and extract image features with different scales. Our experimental results demonstrate that the proposed method significantly improves the reorientation and segmentation performance. This joint learning framework promotes mutual enhancement between reorientation and segmentation tasks, leading to cutting edge performance and an efficient image processing workflow. The proposed end-to-end deep network has the potential to reduce the burden of manual delineation for cardiac images, thereby providing multimodal quantitative analysis assistance for physicists.
\end{abstract}

\keywords{Cardiac SPECT \and Cardiac PET \and deep learning \and LV reorientation \and LV segmentation}

\section{Introduction}
Myocardial perfusion imaging (MPI) plays a critical role in diagnosing coronary artery disease (CAD) \cite{b1}. 
The nuclear cardiac imaging, including single photon emission computed tomography (SPECT) and positron emission computed tomography (PET), are two imaging techniques commonly used in MPI. 
SPECT is known for its cost-effectiveness with tracers and scanners, as well as its flexibility in performing MPI \cite{b2}.
On the other hand, PET offers better spatial resolution and signal-to-noise ratio (SNR). 
By administering radiolabeled tracers and capturing images of the heart at rest and during stress, MPI provides functional information on the segmental heart muscle contraction in the left ventricular (LV) region \cite{b3}. 
The functional indices derived from these imaging biomarkers can accurately evaluate the myocardial blood flow and cardiac function, which in turn helps to assess the CAD risk.

\begin{figure*}
    \centering
    \includegraphics[width=0.8\textwidth]{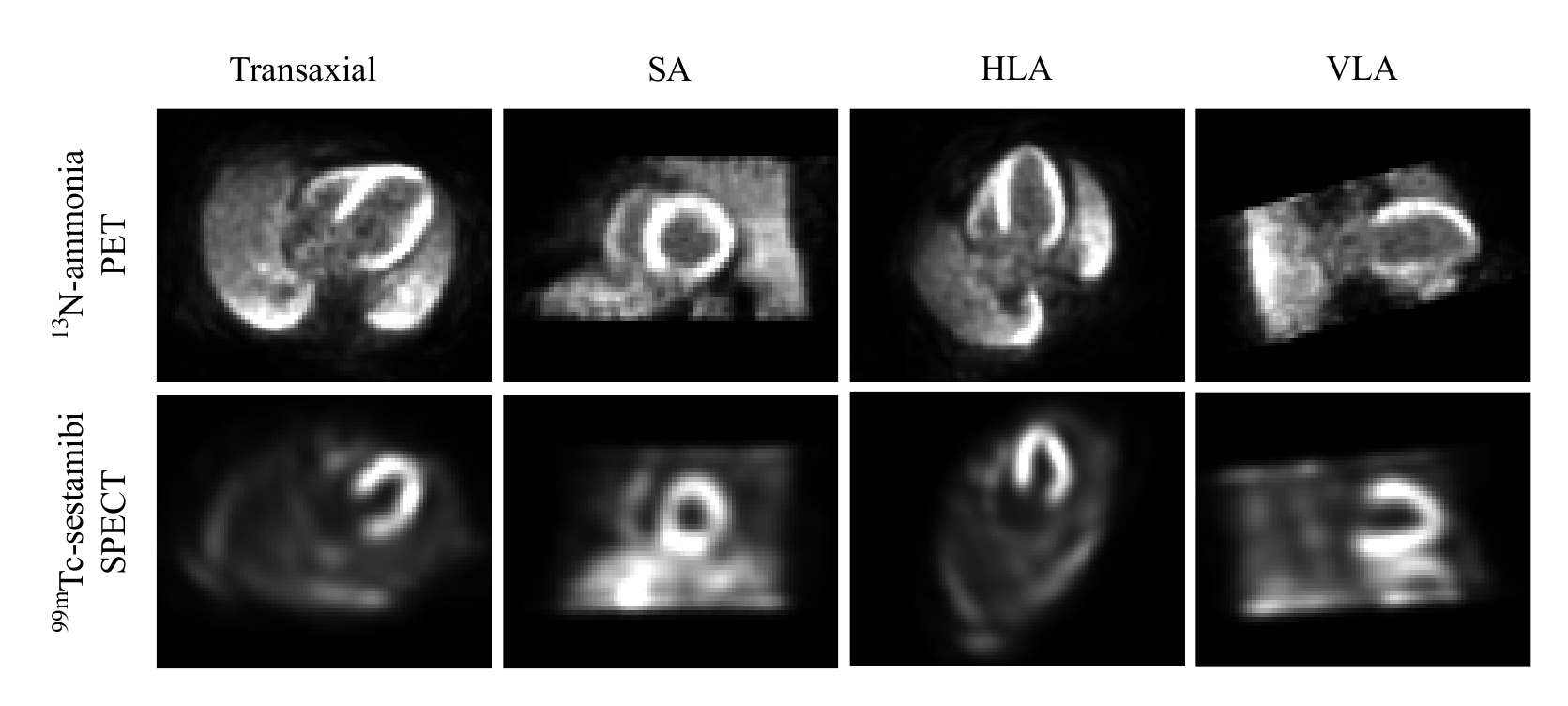}
    \caption{Illustration of the transaxial PET and SPECT images in the first column. The second, third, and fourth colum are the SA, HLA, and VLA slices reoriented by manual method.}
    \label{fig:fig1}
\end{figure*}

However, the long-axis of the heart is not parallel to the axial direction of reconstructed nuclear medical images.
Directly using these transaxial images for LV function assessment may result in numeric errors and even misinterpretation of the cardiac status. 
For precise LV quantitative information, the reconstructed cardiac SPECT/PET images need to be reoriented to specific standard slice-planes, namely short-axis (SA), horizontal long-axis (HLA), and vertical long-axial (VLA) slices, which are illustrated in Fig. \ref{fig:fig1} (\cite{b4}). 
These SA slices, which are perpendicular to the LV’s long-axis, are used in clinical practice to extract quantitative perfusion parameters \cite{b5}. 
The accuracy of reorientation process significantly affects the analysis of the LV image. 

Typically, the reorientation of MPI requires manual detection of the LV’s long-axis in the standard transaxial and sagittal plane \cite{b6}. 
The transformation of coordinates is determined by the long axes of LV in different image planes. However, this procedure is time-consuming and may vary significantly among physicians. 
Therefore, automatic techniques aimed at reorientation are essential for cardiac image analysis.

Image processing of cardiac clinical diagnosis comprises the segmentation of LV myocardium (LV-MY) and the LV blood-pool (LV-BP). 
The quantitative measurement of LV-MY and LV-BP volume is critical for the analysis of cardiac parameters, including end-diastolic volume (EDV), end-systolic volume (ESV), and ejection fraction (EF) \cite{b7}. 

Nevertheless, the segmentation of 3D myocardial perfusion imaging data is difficult. 
Multivariate analysis techniques, such as atlas-based analysis \cite{b8}, threshold-based analysis, cluster analysis \cite{b9}, and independent component analysis \cite{b10}, can aid in the automatic separation of cardiac components and LV-MY and LV-BP segmentation. 
However, these manual-feature-based techniques may not obtain accurate edge information of the LV structures. 
Learning-based feature extraction strategies can enhance the model's generalization capacity and robustness.

Previous studies usually developed reorientation and segmentation algorithms separately \cite{b11,b12}.
However, the target of reorientation operation is highly correlated with the LV-MY and LV-BP segmentation. 
The segmentation network can obtain a mask image that subtly corrects the reorientation parameters. Moreover, utilizing SA slices can enhance the precision of LV-MY segmentation prediction by aligning the cardiac region in the center of the image volume. 
Consequently, we propose an end-to-end cardiac reorientation and segmentation model to effectively accomplish the task of cardiac reorientation and segmentation.

End-to-end LV reorientation and segmentation systems face several challenges. 
Traditional manual reorientation methods rely heavily on overall structure integrity in cardiac images \cite{b13}. 
Due to the structural differences of each patient’s heart, the reorientation method must be able to precisely detect and extract the LV features from the whole nuclear cardiac images. 
The segmentation task faces difficulties in segmenting the small-size LV-MY target in large-scale PET/SPECT images. 
Compared to modalities like magnetic resonance imaging (MRI), PET/SPECT images are harder to semantically segment due to its coarser resolution, lower SNR, and more significant partial volume effect (PVE). 
Further, the different spatial resolutions and image contrasts of PET and SPECT images pose challenges to the feature encoder of the deep learning-based models. 

In this paper, a novel deep-learning-based multi-scale spatial transformer U-Net (MS-ST-UNet) is introduced to address the above issues. 
The proposed end-to-end framework combines a multi-scale spatial transformer network (MSSTN) and a multi-scale UNet (MSUNet) \cite{b14} with the goal of conducting LV reorientation and segmentation jointly. 
MSSTN is designed to accommodate the 3 degrees of translation and the 3 degrees of rotation parameters. The multi-scale outputs are then segmented by the MSUNet. 
The scale transformer (ST) blocks are created to adapt the multi-scale outputs of MSSTN to fit the UNet structure. 
Reorienting the LV region to the image volume center significantly improves segmentation accuracy and efficiency. 
Segmentation network can also subtly correct the reorientation parameters during the procedure. Data from two modalities, $^{13}$N-ammonia PET and $^{99m}$Tc-sestamibi SPECT, are utilized for training and evaluation. 
A series of experimental results demonstrate that the model has achieved state-of-the-art performance in reorientation and segmentation when applied to multi-modal data. 
The contributions of this work are as follows:

\begin{enumerate}
\item This study proposes an automatic end-to-end model, named MS-ST-UNet, which is designed to reorient and segment the LV region from nuclear cardiac images. 
The MSSTN module is responsible for predicting the 6 rigid-body registration parameters, while the MSUNet module produces the LV myocardial and LV blood-pool masks. 
The two components of MS-ST-UNet are connected via a multi-scale sampler and threshold loss. 
The all-in-one model simplifies the quantitative analysis process for MPI data.
\item The MS-ST-UNet employs a multi-scale strategy to improve the accuracy of orientation and segmentation. 
The model includes multi-scale samplers that generates images of different resolutions, and scale transformer (ST) blocks which are designed to align the feature scales propagated by the multi-scale samplers in MSSTN. 
The multi-scale structure integrates features of various scales, enabling the model to better focus on the LV region.
\item This study employed two distinct imaging modalities to assess the model's generalization capacity and robustness. 
The experimental results show that the proposed model demonstrated excellent reorientation and segmentation results on both $^{13}$N-ammonia PET and $^{99m}$Tc-sestamibi SPECT images.
\end{enumerate}

\section{Related Works}
\subsection{LV reorientation}
Various workflows have been developed as substitutes for the manual cardiac reorientation tasks \cite{b15}. 
Researches in this field can be divided into two primary categories: 1) methods based on geometric image features and 2) methods based on deep learning. Mullick et al. developed topological geometry methods to reorient the segmented LV volumes \cite{b11}.
The threshold-based approach is used to extract the 3D LV model from the SPECT images, which incorporated prior knowledge of cardiac anatomy information. 
Slomka et al. proposed a method for registering raw SPECT data to a reoriented cardiac template \cite{b12}.
However, prior knowledge-based methods may not accurately predict the rotation and translation parameters due to the low intensity and structure disparities in some cardiac volumes with myocardial ischemia. 
Deep-learning-based methods have recently been widely adopted for medical image registration. 
These learning-based features have shown great potential in cardiac reorientation task. 
Vigneault et al. proposed the $\Omega$-net (omega-net) to segment the LV-MY and four cardiac chambers from MRI images \cite{b16}.
Without prior information, the $\Omega$-net is capable of outputting reoriented cardiac images. 
Its end-to-end scheme improves training and prediction efficiency. 
However, due to its complex combination of several UNet structure and spatial transform network (STN), the $\Omega$-net is not suitable for use with 3D PET/SPECT data. 
To address this, Zhang et al. developed a CNN based model to reorient the SPECT LV images \cite{b17}.
The STN structure learns the 6 rotation and translation parameters. 
In addition, threshold images are adopted as one of the registration losses. 
Nonetheless, this method cannot handle cardiac data with different scales, as the absence of scaling parameters in this research.

\subsection{Cardiac segmentation}
The segmentation of LV-MY and LV-BP from different modality of images has attracted extensive attention from researchers \cite{b16,b18,b19,b20}. 
Several grand challenges have been set up to achieve more accurate cardiac segmentation methods \cite{b21}. 
Due to the powerful feature extraction capacity, deep learning-based method, especially UNet-based convolutional neural networks, are widely applied for the cardiac segmentation tasks. 
Wang et al. designed a V-Net structure-based network to segment the epicardial and endocardial boundaries from SPECT images \cite{b18}. 
A compound loss function and accurate myocardium volume measurement method are presented in this work. Khened et al. proposed a multi-scale residual network based on densely connected convolutional neural networks (DenseNets) \cite{b19}. 
Compared with the V-Net structure, this work incorporates multi-scale processing in the initial layer by performing convolutions on the input with different kernel sizes. 
This operation, alongside with the long-skip short-skip network structure, yields higher segmentation accuracy for different scale cardiac MRI images. 
The nn-UNet uses a fully-automated dynamic adaptation of the segmentation pipeline to independently segment each task of in Medical Segmentation Decathlon (MSD) \cite{b23,b24}. 
The cascade U-Net structure performs well in different modalities of medical images and for different segment target. 
To construct a precise reorientation and segmentation method, attention must be paid to the impact of cardiac volume scale in PET/SPECT images.

\section{Method}
Due to the relatively small proportion of the heart region in the entire PET/SPECT image, detecting the LV target first and reorienting it to standard planes will improve the network’s segmentation capability. 
Moreover, precise segmentation results will also help the model to obtain a more accurate reorientation prediction. 
Therefore, we propose the MS-ST-UNet, which adopts the stepwise strategy for reorientation and segmentation of cardiac PET/SPECT images in an end-to-end differentiable CNN framework. 
The proposed network comprises two parts: the MSSTN and MSUNet. 
First, the input image $\mathcal{I}$ undergoes the MSSTN module. 
Then, the different output streams of MSSTN with different scaled images are segmented by the MSUNet. 
In the remaining part of this section, we provide a detailed introduction to the model structure and learning framework.

\subsection{Architecture of MS-ST-UNet}
\subsubsection{MSSTN}
\begin{figure*}
    \centering
    \includegraphics[width=0.85\textwidth]{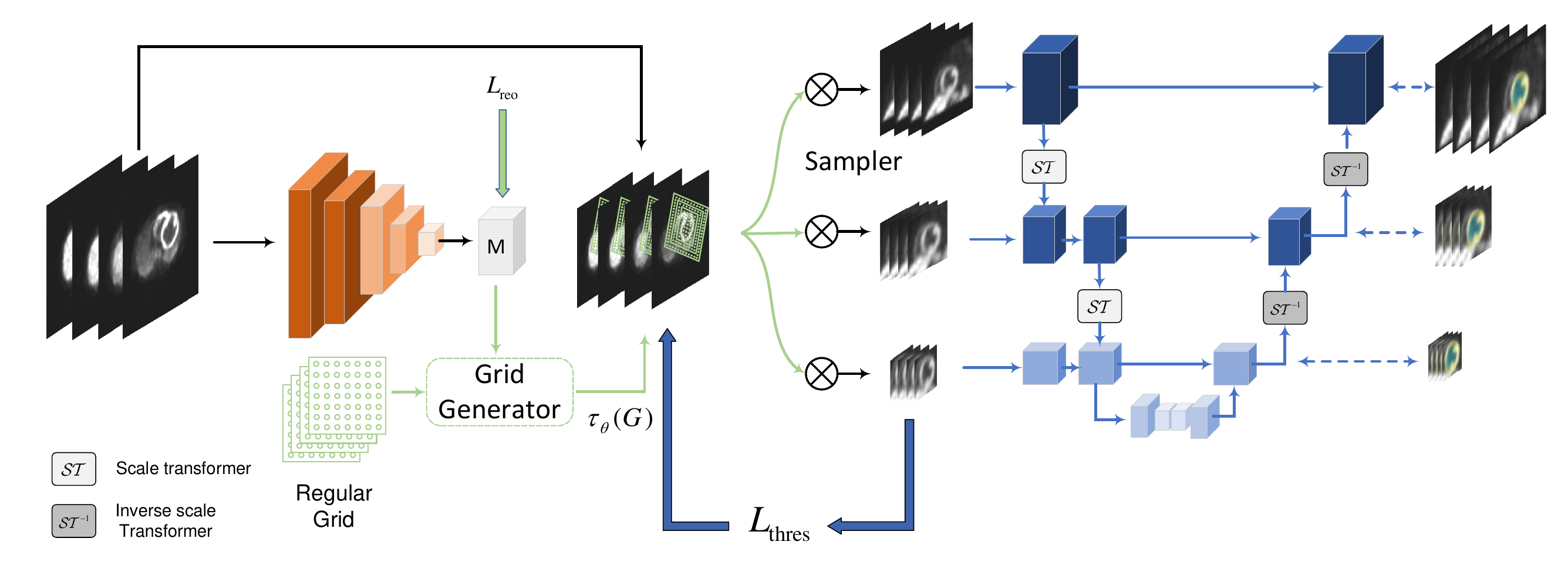}
    \caption{The model structure of MS-ST-UNet. The MSSTN concatenates to the MSUNet to form the MS-ST-UNet.}
    \label{fig:fig2}
\end{figure*}

The spatial transformer network (STN) is a general layer designed to achieve translation invariance in deep models. 
The STN module consists of 3 submodules: a localization network, a parameterized sampling grid generator, and a differentiable image sampling operation. 
First, the localization network takes the input feature map $\boldsymbol{U}$ and generates a affine transformation matrix $\boldsymbol{M}$. 
Then, the grid generator $\tau_{M}$ implements the transformation matrix on the regular feature sampling grid $\boldsymbol{G}$, producing the deformation grid $\boldsymbol{G}{'}$. 
Finally, the deformation grid $\boldsymbol{G}{'}$ is applied to the feature map $\boldsymbol{U}$ to obtain a spatial invariance feature $\boldsymbol{V}$.

We aim to acquire the transformation matrix $\boldsymbol{M}$ from the original transaxial images and subsequently sample the input image into the standard SA image. 
The transformation matrix $\boldsymbol{M}$ can be decomposed into three different types of motion matrices: the rotation matrix $\boldsymbol{R}$, the translation matrix $\boldsymbol{T}$, and the scaling matrix $\boldsymbol{S}$. 
Since the 3D data format, the rotation matrix $\boldsymbol{R}$ necessitates the computation of 3 Euler angles $\theta_{\alpha}$,$\theta_{\beta}$,$\theta_{\gamma}$, which are ordered as YXZ axial direction. 
The rotation matrix can be derived as:
\begin{equation}
\begin{aligned}
& \boldsymbol{R}(\theta)= \\
& \begin{bmatrix} \cos{\theta_{\alpha}} & 0 & \sin{\theta_{\alpha}} \\ 0 & 1 & 0 \\ -\sin{\theta_{\alpha}} & 0 & \cos{\theta_{\alpha}} \end{bmatrix} 
\begin{bmatrix} 1 & 0 & 0 \\ 0 & \cos{\theta_{\beta}} & -\sin{\theta_{\beta}} \\ 0 & \sin{\theta_{\beta}} & \cos{\theta_{\beta}} \end{bmatrix}
\begin{bmatrix} \cos{\theta_{\gamma}} & -\sin{\theta_{\gamma}} & 0 \\ \sin{\theta_{\gamma}} & \cos{\theta_{\gamma}} & 0 \\ 0 & 0 & 1 \end{bmatrix}
\end{aligned}
\label{eq:eq1}
\end{equation}

Following the rotation operation, we apply the translation matrix $\boldsymbol{T}$, which comprises parameters $[t_{x}, t_{y}, t_{z}]$, to shift the LV to the output’s center position. The scaling matrix $\boldsymbol{S}$, which is composed of parameter $s$, is able to alter the output’s scale. The transformation matrix $\boldsymbol{M}$ can be calculated as: 
\begin{equation}
\boldsymbol{M}=[{\boldsymbol{S}}\cdot{\boldsymbol{R}},T^{\rm T}]
\label{eq:eq2}
\end{equation}
Where where the scaling matrix $\boldsymbol{S}$ multiplies the rotation matrix $\boldsymbol{R}$, and the ${*}^{\rm T}$ denotes the matrix transpose operation. $\boldsymbol{S}$ and $\boldsymbol{T}$ can be described as:
\begin{equation}
\boldsymbol{S}(s)=\begin{bmatrix} s & 0 & 0 \\ 0 & s & 0 \\ 0 & 0 & s \end{bmatrix}
\label{eq:eq3}
\end{equation}

\begin{equation}
\boldsymbol{T}(t)=\begin{bmatrix} t_{x} & t_{y} & t_{z} \end{bmatrix}
\label{eq:eq4}
\end{equation}
In Fig. \ref{fig:fig2}, we can observe that the MSSTN and MSUNet employ 3 distinct image scales. 
The use of multi-scale images allows the modules to capture fine details and high-level features, leading to better feature representation. 
The parameter s in scaling matrix $\boldsymbol{S}$ controls the shape of the output. 
In our application, the sampler takes the input image $\mathcal{I}\in {\mathbb{R}^{{C}\times{D}\times{W}\times{H}}}$ ($C=1$, $D=128$, $W=128$, $H=128$) using bilinear interpolation.
We set s to 1.0, 0.5, and 0.25 for the 3 samplers.
At the  $(c,d',w',h')$ location, the output $\mathcal{I}'_{c,d',w',h'}$ is the sum of the input values $\mathcal{I}_{c,d,w,h}$ in the nearest pixel:
\begin{equation}\label{eq:eq5} 
\begin{aligned}
   \mathcal{I}'_{c,d',w',h'} = &\sum_{d=1}^D \sum_{w=1}^W \sum_{h=1}^H \mathcal{I}_{c,d,w,h} \\
   &* max(0,1- \left| (G'_{1,d',w',h'}+1)*\frac{D-1}{2}-d  \right|
   \\
   &* max(0,1- \left| (G'_{2,d',w',h'}+1)*\frac{W-1}{2}-w  \right|
   \\
   &* max(0,1- \left| (G'_{3,d',w',h'}+1)*\frac{H-1}{2}-h  \right|
\end{aligned}
\end{equation}
The grid transform generator is calculated as $\tau_{M}(*)= \boldsymbol{M} \cdot {*} $, and $\boldsymbol{M}$ is shown in Eq. \ref{eq:eq2}. 
The transformed grid $\boldsymbol{G}' \in {\mathbb{R}^{3\times{D}\times{W}\times{H}}}$ is obtained by applying  $\tau_{M}(*)$ to the original grid $\boldsymbol{G} \in {\mathbb{R}^{3\times{D}\times{W}\times{H}}}$ ($\boldsymbol{G}'=\tau_{M}(\boldsymbol{G})$). 
Notably, the differentiability of the sampler enables the gradients to propagate not only to the input but also to the sampling grid (\cite{b25}).
As a result, the model can be trained end-to-end.

\subsubsection{MSUNet}
Multi-scale segmentation models have been developed to improve performance by capturing finer details at different image scales \cite{b28,b29}. 
In this work, we leverage 3 different image scales to enhance the model’s segmentation capacity. 
The outputs of the MSSTN samplers have varying shapes and attention features. 
By feeding these multi-scale cardiac features into the MSUNet’s encoder, we improve the network’s robustness. 
Unlike mainstream multi-scale models that generate multi-scale images by repeatedly downsampling the input image, this approach benefits from the interpolation sampling from MSSTN. 
The multi-scale images generated by the samplers can more prominently highlight the target LV regions, while eliminating the need for multiple downsampling processes and improving the efficiency of model training. 
Due to the interpolation in the MSSTN sampler, traditional pooling method cannot match the features propagated by the reorientation module in terms of feature scale. 
Therefore, we design a scale transformer block to replace the pooling operation.
Inspired by the differentiable sampler in MSSTN, the encoder’s feature $\boldsymbol{F} \in {\mathbb{R}^{{C_f}\times{D_f}\times{W_f}\times{H_f}}}$ has a coordinate grid $\boldsymbol{G}_f \in {\mathbb{R}^{{3\times{D_f}\times{W_f}\times{H_f}}}}$ of pixels $\boldsymbol{G}_{f,i}=(x_{f,i}, y_{f,i}, z_{f,i})$ for each channel ($i=1,2,3$). 
The scale transform block uses scale operation $\tau_s(*)$ to transform the pixel $\boldsymbol{G}_{f,i}$ to the target position $\boldsymbol{G}^t_{f,i}=(x^t_{f,i}, y^t_{f,i}, z^t_{f,i})$. 
This process can be described as:

\begin{equation}
\left(
    \begin{array}{c}
         x^t_{f,i}\\
         y^t_{f,i}\\
         z^t_{f,i}\\
    \end{array} 
\right) = \tau_s(\boldsymbol{G}_{f,i})=[\boldsymbol{S}(s)\boldsymbol{R}(\Delta\theta),\boldsymbol{T}(\Delta t)^{\rm T}]\cdot \left(
    \begin{array}{c}
         x_{f,i}\\
         y_{f,i}\\
         z_{f,i}\\
         1\\
    \end{array} 
\right)
\label{eq:eq6}
\end{equation}

where $s$ is the scaling parameter, $\Delta\theta=\theta_{1}-\theta_{2}$ and $\Delta t=t_1-t_2$ are differences of rotation angles and translation parameters from two scales. 
After obtaining the $\boldsymbol{G}^t_{f,i}$, we can use Eq. \ref{eq:eq5} to get $\boldsymbol{F}' \in {\mathbb{R}^{{{C'_f}\times{D'_f}\times{W'_f}\times{H'_f}}}}$. 
The modified parameter s enables the feature maps to undergo different scaling transforms, addressing the limitation of the pooling operation that cannot accommodate arbitrary input sizes. 
The inverse scale transform blocks use $1/s$ as the scaling parameter, and $\Delta\theta$, $\Delta t$ are all 0.
Additionally, we employ the deep supervision strategy to enhance the prediction accuracy. 
Multiple auxiliary supervision branches are incorporated into the network to enable the learning of complex features at various levels of abstraction.

The MSUNet is connected to the MSSTN, and both modules are trained jointly using loss functions. 
The weights and parameters of both modules are updated simultaneously during training to enable efficient collaboration between the two different operations.

\subsection{Loss function}
We proposed a novel compound loss function that combines the loss of 6 rigid-body parameters with Dice loss for accurate reorientation and segmentation prediction. 
Specifically, our localization network learns to predict rotation and translation parameters, represented as $\boldsymbol{\Theta}=\{{\theta_{\alpha},\theta_{\beta},\theta_{\gamma}}\}$ and $\boldsymbol{T}=\{{t_x,t_y,t_z}\}$.
To achieve this, we use L1 loss to measure the difference between the ground truth and predictions. 
Due to the issue of error amplification between synonymous rotation angles (such as $\pi$ and $-\pi$), we should constrain the rotation angular difference to a range of $-\pi$ to $+\pi$.
Consequently, we define our rotation loss ($L_{rot}$):
\begin{equation}
L_{\rm rot}=\frac{1}{3}\sum_{i}\left| \Omega(\theta_i-\hat{\theta}_i) \right|
\label{eq:eq7}
\end{equation}
where $\Omega(*)=\rm {mod}(*+\pi,2\pi)-\pi$; $\theta_i$ and $\hat{\theta}_i$ ($i=\alpha,\beta,\gamma$) are the ground truth and predictions of rotation angle \cite{b16}. 
Similar to the rotation loss, translation loss can be expressed as:
\begin{equation}
L_{\rm tran}=\frac{1}{3}\sum_{k}\left| t_k-\hat{t}_k \right|
\label{eq:eq8}
\end{equation}
The regression loss $L_{\rm reo}$ of MSSTN is:
\begin{equation}
L_{\rm reo}=\mu L_{\rm rot}+L_{\rm tran}
\label{eq:eq9}
\end{equation}
where $\mu$ is an hyperparameter of the regression loss.
We set $\mu$ as 10 for the balance of rotation and translation loss.

Fig. \ref{fig:fig2} displays the use of a threshold loss to improve the optimization of the MSSTN.
Specifically, this loss constrains the regression output to enhance performance. 
To provide a more intuitive demonstration of the reorientation capabilities, we utilize binarized input images and pass them through the reorientation module with predicted and ground truth parameters, resulting in the calculation of the threshold loss. 
Binarization of input image is performed using its mean intensity. We apply the Dice loss to quantify the dissimilarities between the predicted binary images and the ground truth:
\begin{equation}
L_{\rm thres}=\rm DL(\boldsymbol{B},\hat{\boldsymbol{B}})=1-\frac{2*|\boldsymbol{B} \cap \hat{\boldsymbol{B}}|}{|\boldsymbol{B}|+|\hat{\boldsymbol{B}}|}
\label{eq:eq10}
\end{equation}
where $\hat{\boldsymbol{B}}$ and $\boldsymbol{B}$ are the predicted binary images and ground truth binary images respectively. 

We use the Dice loss as the segmentation loss. 
To make the model more sensitivity to the LV-MY region, hyperparameter $\eta$ is applied to the Dice loss. The segmentation loss $L_{seg}$ is:
\begin{equation}
L_{\rm seg}=\eta \rm DL(\boldsymbol{R}_{\rm MY},\hat{\boldsymbol{R}}_{\rm MY})+\rm DL(\boldsymbol{R}_{\rm BP},\hat{\boldsymbol{R}}_{\rm BP})+\rm DL(\boldsymbol{R}_{\rm BG},\hat{\boldsymbol{R}}_{\rm BG})
\label{eq:eq11}
\end{equation}
where $\boldsymbol{R}_{\rm MY}$, $\boldsymbol{R}_{\rm BP}$, $\boldsymbol{R}_{\rm BG}$ are the ground truth region of LV-MY, LV-BP, and background, respectively; $\hat{\boldsymbol{R}}_{\rm MY}$, $\hat{\boldsymbol{R}}_{\rm BP}$, $\hat{\boldsymbol{R}}_{\rm BG}$ are the predicted LV-MY, LV-BP, and background area, respectively. $\eta$ is set as 3 in this study.
Three parts of losses are summed together to obtain the final loss:
\begin{equation}
L=L_{\rm reo}+L_{\rm thres}+L_{\rm seg}
\label{eq:eq12}
\end{equation}

\section{Experiments}
\subsection{Data preparation}
\begin{table*}[!htbp]
\centering
\captionsetup{width=\linewidth,labelformat=default,labelfont=bf, font=normalsize}
\caption{Acquisition information of the PET and SPECT dataset}
\label{tab:tab1} 
\small 
\renewcommand{\arraystretch}{1.5}
\setlength{\tabcolsep}{10pt}
\begin{tabular}{l l l}
\toprule[1pt]
                         & Center A (PET)   & Center B (SPECT)  \\
\midrule
Gender (male/female)     & 75/10            & 42/18             \\
Age (years)              & $58.59 \pm 9.86$ & $57.25 \pm 13.40$ \\
Manufacturer             & SIEMENS          & Philips                \\
Slice thickness (mm)     & 5                & 6.80              \\
Pixel spacing (mm/Pixel) & 2.03             & 6.80              \\
Energy Window (keV)      & [425, 650]       & [130, 151]        \\
Radiopharmaceutical      & $^{13}$N-ammonia & $^{99m}$Tc-sestamibi \\
Injected dose (Mbq)            & 323 to 999    & 740 to 1332    \\
Reconstruction Method    & OSEM3D           & OSEM3D           \\
\bottomrule[1pt]
\end{tabular}
\end{table*}

In this study, we used two distinct in-house datasets to evaluate the performance of MS-ST-UNet: a $^{13}$N-ammonia PET dataset from center A and a $^{99m}$Tc-sestamibi SPECT dataset from center B.
The different imaging modality and reconstruction parameters across these two datasets posed a challenge for MS-STN-UNet to achieve higher levels of generalization. 
The acquisition information of the two datasets are shown in Table 1. All the PET and SPECT data were reconstructed with a size of $128 \times 128 \times 128$. For each cardiac PET/SPECT image, patient anonymity was ensured and written consent was obtained prior to the commencement of the study.

\begin{table*}[!htbp]
\centering
\captionsetup{width=0.6\linewidth,labelformat=default,labelfont=bf, font=normalsize}
\caption{Statistical information for the 6 rigid registration parameters of PET and SPECT dataset. The data is presented as \textit{PET/SPECT}.}
\label{tab:tab2} 
\small 
\renewcommand{\arraystretch}{1.5}
\setlength{\tabcolsep}{10pt}
\begin{tabular}{lllll}
\toprule[1pt]
\textit{PET/SPECT}       & Min   & Max  & Mean   & Std  \\
\midrule
$\theta_{\alpha}$  & -1.21/-1.49 & -0.34/-0.26 & -0.79/-0.83 & 0.18/0.22 \\
$\theta_{\beta}$   & -1.69/-1.68 & -0.48/-0.32 & -0.97/-1.14 & 0.21/0.27 \\
$\theta_{\gamma}$  & 0.56/0.66 & 1.67/1.80 & 1.21/1.22 & 0.19/0.27 \\
$t_x$ & -0.28/-0.30 & -0.04/-0.06 & -0.18/-0.19 & 0.05/0.06 \\
$t_y$ & -0.02/-0.06 & 0.22/0.31 & 0.10/0.15 & 0.06/0.08 \\
$t_z$ & -0.33/-0.47 & -0.17/-0.04 & -0.24/-0.29 & 0.03/0.09 \\
\bottomrule[1pt]
\end{tabular}
\end{table*}

\subsection{Center A PET data}
The PET dataset used in this study was collected from Guangdong General Hospital and contains 85 anonymized patients. 
To create a reliable model for PET image analysis, we divided the dataset into two sets: a training set of 71 cases and a test set of 14 cases. 
All PET scans were reconstructed and manually reoriented by an experienced medical physicist to ensure consistency and accuracy. 
We then used the IDL software provided by University of Massachusetts Medical School \cite{b30} to obtain the 6 rigid registration parameters between the transaxial and SA images. 
These parameters were divided into two categories: the rotation parameter $\boldsymbol{\Theta}=\{\theta_{\alpha},\theta_{\beta},\theta_{\gamma}\}$ and the translation parameter $\boldsymbol{T}=\{t_x,t_y,t_z\}$.
The minimum, maximum, mean, and stand deviation of the registration parameters are shown in Table \ref{tab:tab2}. 
To ensure the accuracy of our segmentation results, we labeled the ground truth with the help of 3 experienced medical physicists. 
One medical physicist conducted the initial LV region delineation. 
Then, these labels were reviewed and revised by the other two medical physicists. 
To enlarge the size of the training set, we performed rotation and shifting operations to augment the PET dataset. 
According to the parameters in Table \ref{tab:tab2}, we randomly generate 10 sets of transformation parameters for each PET image. 
We than applied the inverse transformation operation using these parameters to generate the augmented data. 
To guarantee that the cardiac information was preserved during this process, we carefully examined each generated image and removed any with damaged cardiac details. 
In total, we generate 538 transformed images are generated, resulting in a total of 609 images for training and validation.

\subsection{Center B SPECT data}
The SPECT dataset used in this study was collected from University of Massachusetts Medical School and comprises 60 patients. 
The labeling process of this dataset is identical to that of the PET dataset. 
For training and testing, we randomly selected 50 and 10 scans, respectively. 
Data augmentation is performed using the same strategy as in the PET dataset, resulting in a total of 492 images in the training set.

\subsection{Implementation details}
In this study, all methods are implemented by PyTorch \cite{b26} on a Ubuntu 18.04 server with an Nvidia Tesla-V100 graphics processing unit (GPUs) and 128 GB memory.
To conduct the experiments, we utilized a polynomial dynamic learning strategy to adjust the learning rate, where the initial learning rate $l_i$ was set to $1e^{-4}$. 
$l_i$ was multiplied by $(1.0-\frac{t}{t_{max}})^{0.9}$ after each iteration t, and the total iteration $t_{max}=20k$.
We used the Adam optimizer (weight decay=$1e^{-4}$, momentum=0.9) and Eq. \ref{eq:eq11} as the loss function. During the training phase, all models were trained using the same hyperparameters, and we set the batch size to 2.

\subsection{Evaluation metrics and experimental setup}
Four common metrics are used for evaluation. The Dice similarity coefficient (DSC) and 95\% Hausdorff distance (HD95) are utilized to evaluate the segmentation task.
DSC is widely applied in medical image segmentation field, and is calculated based on the precision and recall of the predicted images. 
The HD is highly sensitive to the contour, while DSC focuses on the internal information. 
For reorientation operation evaluation, mean square error (Rot-MSE) and linear regression analysis are applied.

This study conducted three experiments to evaluate the performance of MS-ST-UNet in reorientation and segmentation. 
First, several models with varying reorientation and segmentation strategies were implemented as baseline models to evaluate the performance of MS-ST-UNet. 
These models include a large-scale ($128 \times 128 \times 128$) segmentation network named as Seg Op1, a multi-scale ($128 \times 128 \times 128$, $64 \times 64 \times 64$, $32 \times 32 \times 32$) segmentation network named as Seg Op2, a large-scale end-to-end reorientation segmentation network named as Rot Op1-Seg Op1, and a separated multi-scale reorientation and segmentation network named as Sep Rot Op2-Seg Op2. 
Additionally, state-of-the-art models, the $\Omega$-Net \cite{b16}, V-Net structure \cite{b18}, Dense-U-Net \cite{b19}, and the nnU-Net \cite{b23}, are used to illustrate the importance of multi-scale strategy in this segmentation task. The implementation of each model is obtained from the GitHub codebase. 
Finally, quantitative measurement of the LV-MY volume is conducted to assess the segmentation results of MS-ST-UNet, Rot Op1-Seg Op1, and Sep Rot Op2-Seg Op2.

\section{Results}
\subsection{Results of different reorientation and segmentation strategies}
\begin{sidewaystable}
\centering
\captionsetup{width=\textwidth,labelformat=default,labelfont=bf, font=normalsize}
\caption{Quantitative reorientation and segmentation results obtained on PET dataset. The results are presented as mean [max, min]. Results marked in \textbf{bold} are the best ones.}
\label{tab:tab3} 
\begin{threeparttable}
\footnotesize 
\renewcommand{\arraystretch}{2}
\setlength{\tabcolsep}{6pt}
\begin{tabular*}{\linewidth}{lllllll}
\toprule[1pt]
\multirow{2}{*}{Methods\tnote{1}} & \multicolumn{3}{l}{PET} & \multicolumn{3}{l}{SPECT}  \\
\cmidrule[1pt](r){2-4} \cmidrule[1pt](r){5-7}
{} & LV-MY (DSC/HD95) & LV-BP (DSC/HD95) & Rot-MSE ($10^{-2}$) & LV-BP (DSC/HD95) & LV-BP (DSC/HD95) & Rot-MSE ($10^{-2}$) \\
\midrule[1pt]
\multirow{2}{*}{Seg Op1 (wo Rot)\tnote{2}} & 81.18 [77.96, 82.27] & 93.48 [92.03, 94.03] & \multirow{2}{*}{--} & 77.05 [75.09, 83.02] & 82.73 [81.74, 86.35] &  \multirow{2}{*}{--} \\
{} & 9.75 [7.20, 10.76] & 4.48 [4.00, 5.64] & {} & 10.33 [8.44, 18.27] & 5.59 [4.53, 8.77] &  {} \\
\Xhline{1pt}
\multirow{2}{*}{Seg Op2 (wo Rot)\tnote{3}} & 85.95 [82.03, 89.36] & 94.52 [90.08, 96.58] & \multirow{2}{*}{--} & 88.72 [88.09, 90.08] & 92.62 [92.06, 93.58] &  \multirow{2}{*}{--} \\
{} & 8.78 [6.12, 1.18] & 4.76 [3.96, 5.64] & {} & 9.91 [8.27, 9.36] & 4.36 [3.27, 5.70] &  {} \\
\Xhline{1pt}
\multirow{2}{*}{Rot Op1-Seg Op1\tnote{4}} & 87.03 [80.42, 91.14] &95.19 [92.33, 96.85] & \multirow{2}{*}{1.53 [1.63, 3.07]} & 89.09 [88.66, 92.15] & 92.62 [92.06, 93.58] &  \multirow{2}{*}{4.27 [1.45, 19.64]} \\
{} & 8.92 [7.22, 10.38] & 6.58 [3.60, 5.60] & {} & 9.03 [7.56, 9.87] & 5.15 [3.22, 6.65] &  {} \\
\Xhline{1pt}
\multirow{2}{*}{\makecell{Sep Rot Op2-\\Seg Op2}\tnote{5}} & 90.03 [83.42, 92.65] & $\boldsymbol{96.88}$ [95.16, 96.91] & \multirow{2}{*}{3.62 [1.57, 4.12]} & 93.83 [87.26, 96.12] & $\boldsymbol{96.26}$ [95.23, 96.84] &  \multirow{2}{*}{6.11 [0.10, 7.42]} \\
{} & $\boldsymbol{8.26}$ [7.06, 9.65] & 4.62 [3.84, 5.12] & {} & 8.61 [5.62, 9.66] & $\boldsymbol{4.10}$ [3.15, 7.21] &  {} \\
\Xhline{1pt}
\multirow{2}{*}{\makecell{Rot Op2-Seg Op2 \\ (MS-ST-UNet)}} & $\boldsymbol{91.48}$  [87.55, 92.46] & 96.70 [95.21, 97.35] & \multirow{2}{*}{$\boldsymbol{1.33}$ [0.22, 3.15]} & $\boldsymbol{94.81}$ [94.51, 95.37] & 96.21 [96.03, 96.51] &  \multirow{2}{*}{$\boldsymbol{1.38} $[0.09, 5.34]} \\
{} & 8.34 [7.48, 10.76] & $\boldsymbol{4.38}$ [3.46, 4.90] & {} & $\boldsymbol{8.30}$ [6.04, 9.57] & 4.24 [3.00, 6.46] &  {} \\
\bottomrule[1pt]
\end{tabular*}

\begin{tablenotes}
        \footnotesize
        \item[1] In each segmentation results, the first and second row are DSC (\%) and HD95 (mm) metrics respectively.
        \item[2] Seg Op1 denotes large-scale ($128 \times 128 \times 128$) LV-MY and LV-BP segmentation.
        \item[3] Seg Op2 denotes multi-scale ($128 \times 128 \times 128$, $64 \times 64 \times 64$, $32 \times 32 \times 32$) LV-MY and LV-BP segmentation.
        \item[4] Rot Op1 stands for single-scale ($128 \times 128 \times 128$) reorientation operation.
        \item[5] Rot Op2 represents multi-scale ($128 \times 128 \times 128$, $64 \times 64 \times 64$, $32 \times 32 \times 32$) reorientation operation.
      \end{tablenotes}
\end{threeparttable}
\end{sidewaystable}

\begin{figure*}
    \centering
    \includegraphics[width=0.85\textwidth]{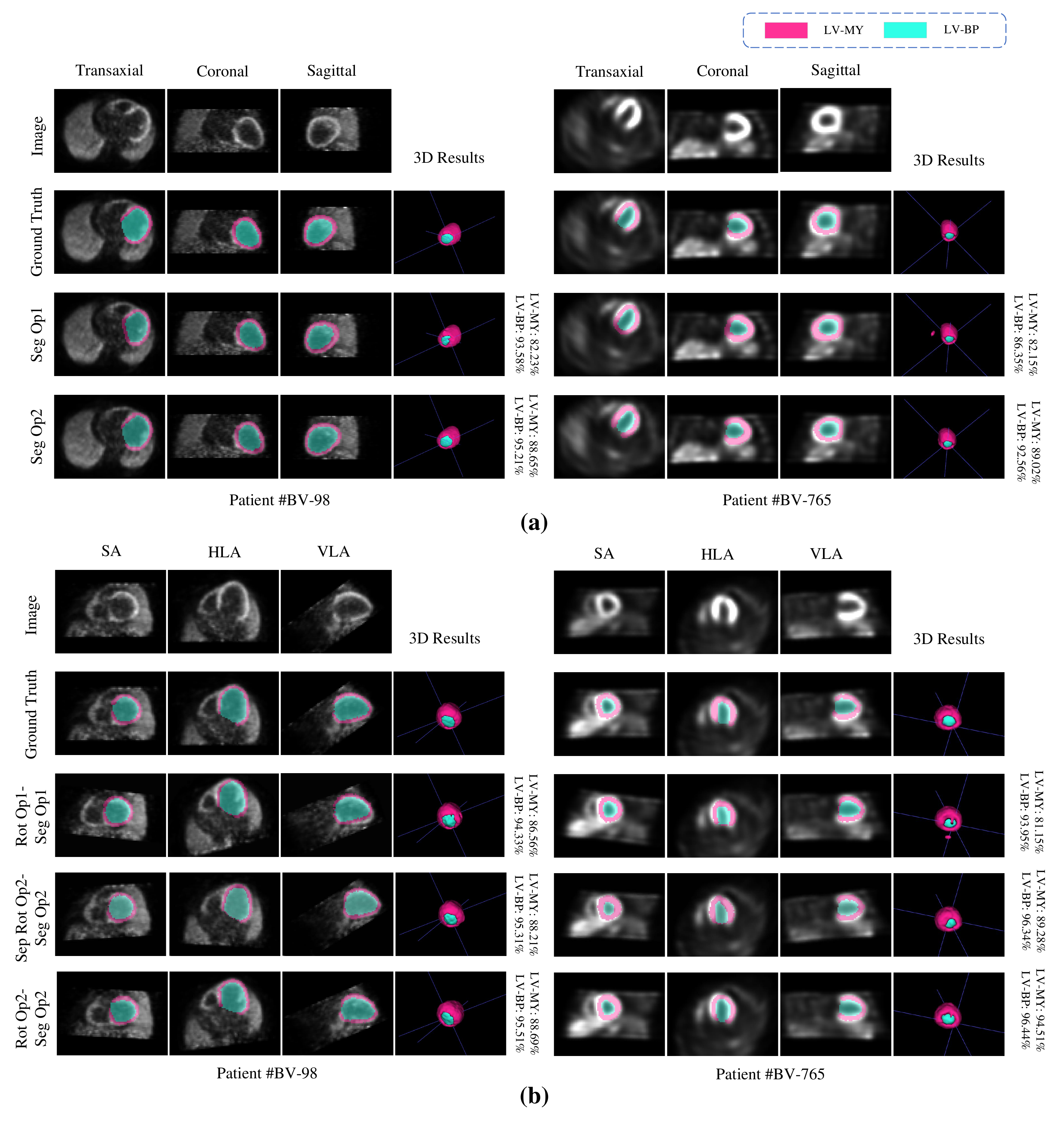}
    \caption{The reorientation and segmentation results of different models. (a) presents the segmentation results of Seg Op1 (wo Rot) and Seg Op2 (wo Rot). (b) represents the reorientation and segmentation output of the Rot Op1-Seg Op1, Rot Op2-Seg Op2, and Manual Rot-Seg Op2.}
    \label{fig:fig3}
\end{figure*}

\begin{figure*}
    \centering
    \includegraphics[width=0.85\textwidth]{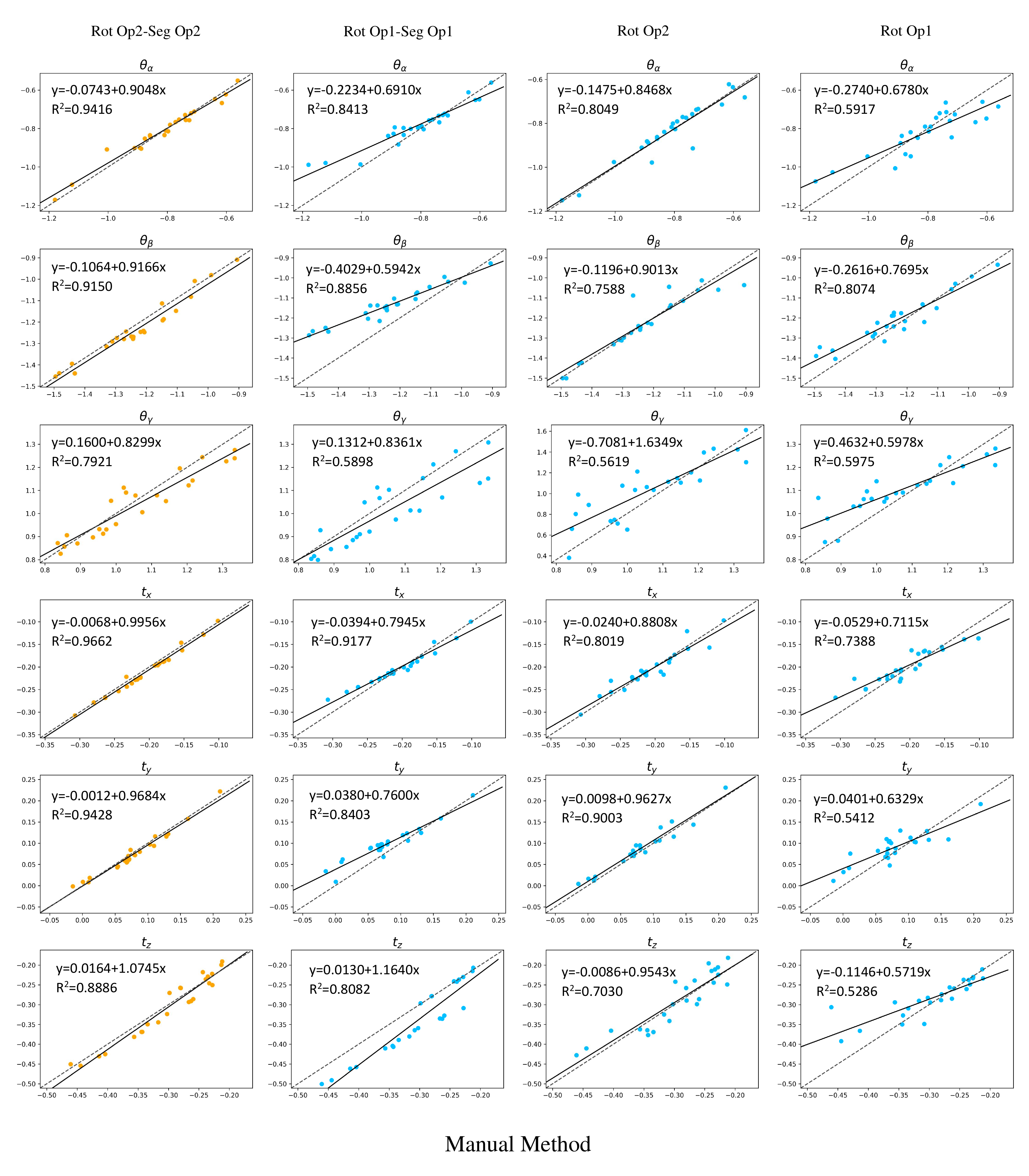}
    \caption{Liner regression analysis result of the predicted 6 rigid-body parameters from the Rot Op2-Seg Op2, Rot Op1-Seg Op1, Rot Op1, and Rot Op2 compared with the manual method.}
    \label{fig:fig4}
\end{figure*}

We assess the performance of MS-ST-UNet across different orientation and segmentation strategies.
ResNet50 and 3D Attention-U-Net \cite{b27} serve as the backbone reorientation and segmentation networks, respectively. 
Table \ref{tab:tab3} presents quantitative results for various reorientation and segmentation methods applied to two cardiac image modalities. 
Segmentation prediction performance is assessed using DSC and HD95 metrics, while Rot-MSE is employed for reorientation estimation.
The results indicate that the multi-scale segmentation (Seg Op2) outperforms large-scale segmentation (Seg Op1) in both PET and SPECT datasets in terms of DSC and HD95 metrics.
Multi-scale reorientation (Rot Op2) also performs better than single-scale (Rot Op1) in both modalities.
Compared to the Sep Rot Op2-Seg Op2 model, the joint learning strategy greatly improves the performance of Rot-MSE.
The segmentation procedure can adjust the reorientation parameters.
Moreover, the MS-ST-UNet achieves the best overall results among all methods, with 91.48\% PET LV-MY DSC, 8.34mm PET LV-MY HD95, 94.81\% SPECT LV-MY DSC, and 8.30mm SPECT LV-MY HD95 metrics.

Fig. \ref{fig:fig3} visually compares the results obtained from the various methods employed in this study.
Specifically, Fig. \ref{fig:fig3}(a) displays the segmentation outcomes of Seg Op1 and Seg Op2, while Fig. \ref{fig:fig3}(b) illustrates the reorientation and segmentation outputs of Rot Op1-Seg Op1, Rot Op2-Seg Op2, and Sep Rot Op2-Seg Op2.
The slice-level and 3D results demonstrate that the proposed multi-scale method has a higher overlap ratio with the ground truth when compared to single-scale models.
The single-scale models exhibit mis-segmentations when predicting SPECT case BV-765.
Although the segmentation results for Rot Op2-Seg Op2 are similar to those of Sep Rot Op2-Seg Op2, the LV-MY prediction of Rot Op2-Seg Op2 shows a higher degree of alignment with the ground truth.
Moreover, the incorporation of the reorientation operation in MS-ST-UNet results in superior LV-MY and LV-BP segmentation performance compared to both Seg Op1 and Seg Op2 models.

Fig. \ref{fig:fig4} presents the linear regression analysis result for all 24 testing cases, with the manual reorientation method as the standard reference.
Notably, the multi-scale reorientation model shows superior predictive performance, reflected in a higher determination coefficient ($\rm R^2$) compared to the single-scale method Rot Op1.
Additionally, the joint learning strategy also improves the reorientation performance. 
According to the Rot-MSE results in Table 3 and the $\rm R^2$ value in Fig. \ref{fig:fig4}, the incorporation of features at varying scales enhances the accuracy of predictions for rigid body parameters, with larger scale features providing a broader context and smaller scale features offering greater detail.
The integration of a multitask structure further assists in fine-tuning the STN structure.

\subsection{Results of different state-of-the-art segmentation methods}
\begin{table*}[!htbp]
\centering
\captionsetup{width=0.85\linewidth,labelformat=default,labelfont=bf, font=normalsize}
\caption{Segmentation results (DSC \%) obtained on the state-of-the-art methods. The results are presented as mean [max, min]. Results marked in \textbf{bold} are the best ones.}
\label{tab:tab4} 
\renewcommand{\arraystretch}{1.5}
\setlength{\tabcolsep}{10pt}
\resizebox{0.85\textwidth}{!}{
\normalsize 
\begin{tabular*}{\linewidth}{lllll}
\toprule[1pt]
Methods       & LV-MY PET   & LV-BP PET  & LV-MY SPECT  & LV-BP SPECT \\
\midrule
$\Omega$-Net  & 87.25 [81.56, 90.66] & 94.91 [91.62, 95.61] & 92.19 [66.23, 94.58] & 92.61 [76.56, 95.78] \\
V-Net   & 88.03 [82.98, 91.05] & 95.25 [93.77, 96.88] & 91.11 [69.41, 96.87] & 94.74 [77.77, 98.13] \\
Dense-UNet  & 88.66 [79.35, 92.88] & 96.00 [72.98, 98.26] & 90.11 [73.16, 94.71] & 91.63 [79.07, 96.25] \\
nn-UNet & 88.19 [81.54, 92.84] & 95.93 [91.89, 97.71] & 91.98 [75.94, 95.21] & 93.99 [88.50, 96.01] \\
MSUNet & \textbf{92.31} [89.55, 93.23] & \textbf{96.23} [95.25, 97.75] & \textbf{93.35} [90.04, 95.34] & \textbf{96.90} [93.54, 97.85] \\
\bottomrule[1pt]
\end{tabular*}
}
\end{table*}

\begin{figure*}
    \centering
    \includegraphics[width=0.75\textwidth]{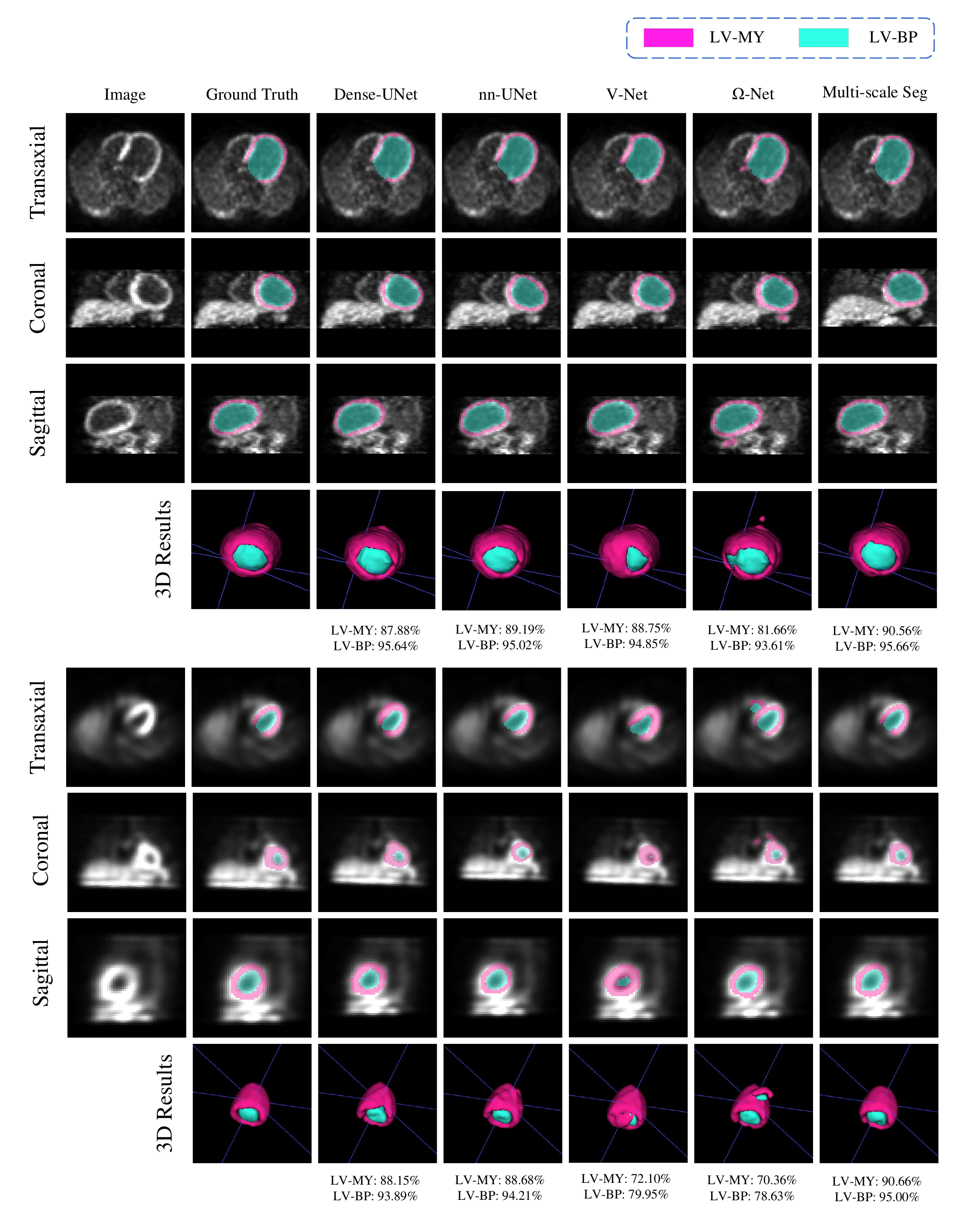}
    \caption{The segmentation results of two PET and SPECT cases by different segmentation methods.}
    \label{fig:fig5}
\end{figure*}

In this part, we conduct experiments to compare the performance of MS-ST-UNet with several state-of-the-art researches.
Table \ref{tab:tab4} represents the performance of each cutting-edge method for segmenting the LV-MY and LV-BP in PET and SPECT images. 
Notably, MSUNet outperforms other models with average improvements of 4.28\%, 0.71\%, 2.00\%, and 3.65\% in terms of DSC for PET LV-MY, PET LV-BP, SPECT LV-MY, and SPECT LV-BP segmentation, respectively. 
Especially for the small-scale LV-MY region, MSUNet outperforms other deep learning models and obtain more precise region delineation. 
Fig. \ref{fig:fig5} shows the segmentation outputs of different models. It is worth noting that the $\Omega$-Net achieves relatively low LV-MY segmentation results with several false-positive LV-MY and LV-BP regions. 
That may due to its complex cascaded UNet structure has great potential of overfitting. 
While the V-Net exhibits over-prediction of the LV-MY region, leading to suboptimal outcomes. 
Both MSUNet and nnUNet demonstrate comparable capabilities in boundary prediction. 
The results obtained by MSUNet achieved the most accurate segmentation results with fewer over-/under-segmentation regions in visual comparison. 

\subsection{Volume measurement of different orientation and segmentation strategies}
\begin{figure*}
    \centering
    \includegraphics[width=0.85\textwidth]{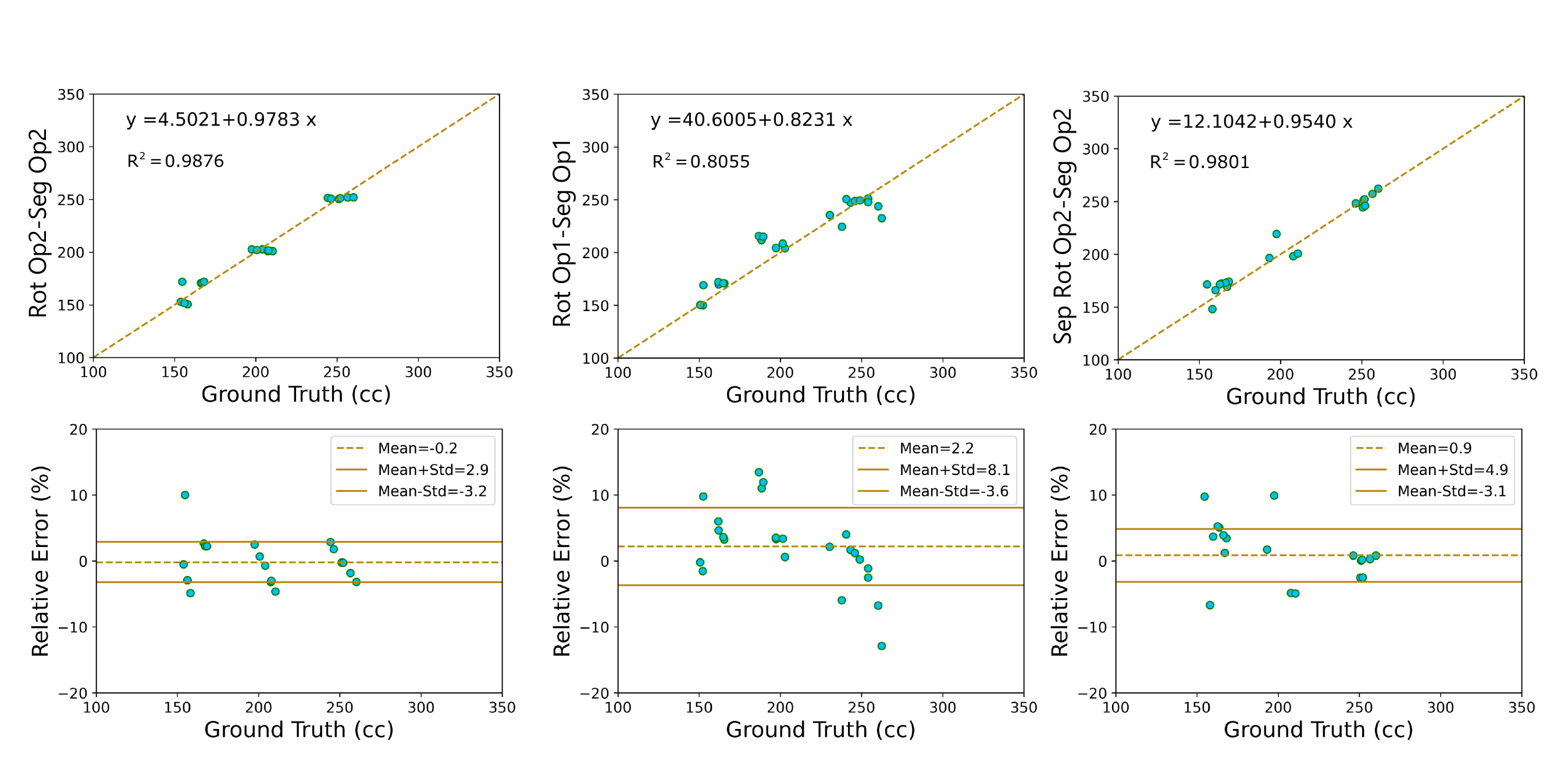}
    \caption{Up: linear regression analysis result of the predicted LV-MY volume with the ground truth. Down: the relative error of the LV-MY volume measured by different models.}
    \label{fig:fig6}
\end{figure*}

\begin{figure*}
    \centering
    \includegraphics[width=0.65\textwidth]{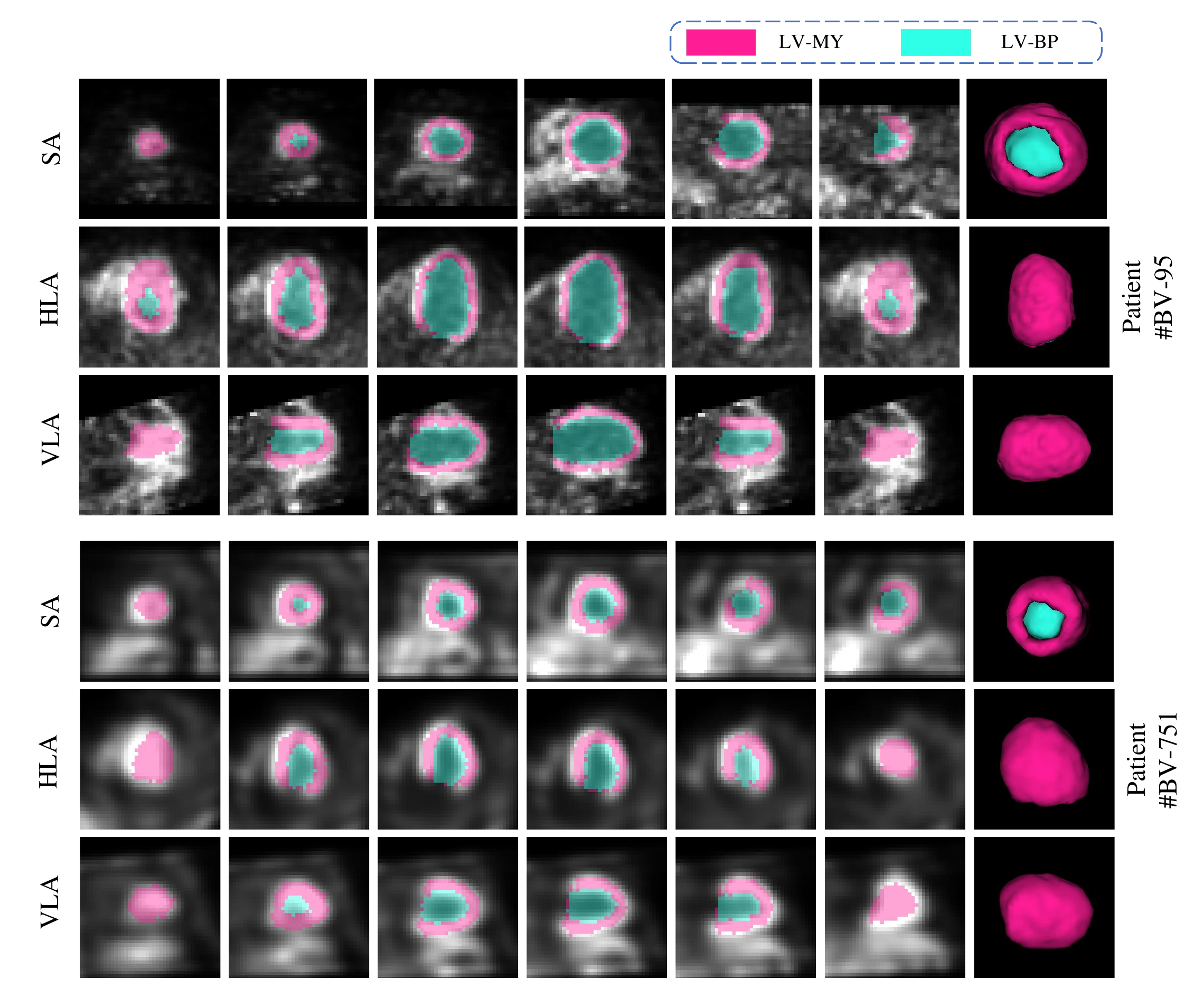}
    \caption{The segmentation results of PET and SPECT data, which consist of multi-layer images, and the 3D reconstructed volume of the LV.}
    \label{fig:fig7}
\end{figure*}

Quantitative measurement of the segmented volume is facilitated by the results of 3D LV segmentation. 
We compare our method, MS-ST-UNet, with the models Rot Op1-Seg Op1 and Sep Rot Op2-Seg Op2. 
Fig. \ref{fig:fig6} shows that the MS-ST-UNet achieved $\rm R^2$ value of 0.9876, which indicates a significant linear correlation between LV-MY volumes measured by proposed method and ground truth. 
The second row of the Fig. \ref{fig:fig6} presents a Bland-Altman plot in LV-MY volumes measurement for each network. 
The MS-ST-UNet outperforms other comparison models, with a mean (and standard deviation) of $-0.2 \pm 3.05$\%. Fig. \ref{fig:fig7} shows the 6 different prediction slices of patient BV-95 (PET) and patient BV-751 (SPECT) generated by MS-ST-UNet. 
The segmentation results provide a comprehensive reconstruction of patient’s current LV structure. 
Some crucial clinical cardiac evaluation parameters, such as EF, EDV, and ESV, could not be evaluated in this work due to the unavailability of appropriate nuclear cardiac images. 
However, the exceptional precision of the model in evaluating LV-MY volumes can potentially enhance the results of further researches in cardiac status assessment.

\section{Discussion}
The MS-ST-UNet model employs a multi-scale approach to connect the MSSTN and MSUNet modules, which guaranteeing precise outcomes in reorientation and segmentation.
This comprehensive framework provides an end-to-end solution encompassing both rigid-body registration and LV segmentation tasks.
The average DSC and HD95 metrics for the contours generated by MS-ST-UNet are larger than 91\% and less than 9.00mm, respectively.
In segmenting small-scale LV-MY region, our method achieves 91.48\% and 84.81\% DSC score for PET and SPECT images, which outperforms the state-of-the-art comparative models. 
The reorientation outcomes of MS-ST-UNet surpass those of both the single-scale end-to-end framework (Rot Op1-Seg Op1) and the separated multi-scale reorientation and segmentation model (Sep Rot Op2-Seg Op2).
These findings provide compelling evidence for the potential of the proposed end-to-end multi-scale framework in achieving accurate reorientation and segmentation outcomes in PET/SPECT image analysis.

By incorporating multi-scale samplers within the MSSTN module and scale transformer blocks within MSUNet, we can effectively address scale variation between reorientation and segmentation operation.
Using scaling and interpolation operations allows the model to concentrate the feature maps on the LV region in nuclear cardiac images.
Fig. \ref{fig:fig4} presents the correlation analysis of the single-scale reorientation and segmentation framework (Rot Op1-Seg Op1), Multi-scale reorientation and segmentation framework (Rot Op2-Seg Op2), single-scale STN (Rot Op1), and multi-scale STN (Rot Op2) with the manual method. 
The utilization of multi-scale techniques and the joint training strategy yielded respective improvements of 0.1074 and 0.1660 in the $\rm R^{2}$ values. 
This underscores the effectiveness of the multi-scale approach, which is tailored to the extraction of LV structure features, leading to improved accuracy in reorientation task. 
Moreover, the segmentation task contributes to refining the performance of the STN.
Table \ref{tab:tab3} shows the competitive results of the MS-ST-UNet. 
The multi-scale strategy greatly boosts the LV-MY segmentation performance, 
achieving an average gain of 5.09\% and 0.66mm in terms of DSC and HD95 metrics. 
The multi-scale samplers and scale transformer blocks in MS-ST-UNet proves to be effective in enhancing the model’s ability to detect and manipulate the LV region in nuclear cardiac images. 
The MS-ST-UNet uses an end-to-end framework to address the LV reorientation and segmentation task. 
A detailed examination of Table 3 reveals that the separated learning framework, Sep Rot Op2-Seg Op2, obtains Rot-MSE values of $3.62 \times 10^{-2}$ and $6.11 \times 10^{-2}$ for PET and SPECT reorientation parameter prediction, respectively. 
Notably, these values are $2.19 \times 10^{-2}$ and $3.29 \times 10^{-2}$ higher than the corresponding averages achieved by joint learning frameworks. 
The LV-MY region delineation by MS-ST-UNet exhibits greater proximity to the ground truth than that of Sep Rot Op2-Seg Op2 from visual comparison in Fig. \ref{fig:fig3}. 
We can conclude that the structure of the joint learning model facilitates mutual promotion between the MSSTN and MSUNet modules, resulting in improved accuracy and efficiency in processing LV reorientation and segmentation tasks in PET and SPECT images.

In this work, we use two different modalities of nuclear cardiac data (PET and SPECT) to train and test the proposed MS-ST-UNet. 
Compared to the pervious automatic methods, our end-to-end reorientation and segmentation model extract multiscale LV structure information and obtain good generalization capacity in multi-modal and multi-center datasets. 
However, there still some limits in the present study. 
The multi-center and multi-modality effects still influence the network performance. 
When applying a model trained on PET dataset to predict SPECT images, the segmentation and reorientation results can only achieve 52\% of the results shown in Table \ref{tab:tab3}. 
To solve this problem, a multi-center effect correction method needs to be applied in the network.
Furthermore, although we used a total of 145 PET and SPECT patients with data augmentation for the network training and testing, additional clinical data from different medical center would further improve the generalizability of the proposed method.

\section{Conclusion}
In this study, we proposed a novel network called MS-ST-UNet for reorienting and segmenting the cardiac LV region. 
The MSSTN module and MSUNet module are connected via a multi-scale sampler and a threshold loss. 
Our multi-scale strategy and scale transformer blocks integrate image features from different scales, helping the model to focus better on the LV area. 
Two different nuclear medical image modalities, $^{13}$N-ammonia PET and $^{99m}$Tc-sestamibi SPECT images, are applied to evaluate the generalization capacity and robustness of the proposed method. 
The extensive experimental results show that the significant improvement brought by the multi-scale strategy. 
Our end-to-end framework can greatly reduce the burden on cardiologists in analyzing MPI. 
The MS-ST-UNet is an ideal candidate for automatically reorienting and quantitatively analyzing cardiac data.

\section*{Acknowledgments}
We express our gratitude to Dr. Michael A. King and Dr. P. Hendrik Pretorius from Radiology Department of University of Massachusetts Medical School, for providing support for this study in terms of clinical datasets as well as support for the software used to achieve manually reorientation of the SPECT cardiac images. 
This work was supported by research grants from National
Natural Science Foundation of China (62101510), National
Natural Science Foundation of China (62001425), and Key Research and
Development Program of Zhejiang Province (2021C03029).

\bibliographystyle{unsrt}  

\end{document}